
\overfullrule=0pt
\magnification 1200
\hsize=6.6truein
\advance\hsize by 2.8truept
\baselineskip 14pt
\tolerance 1000

\def\cite#1{[#1]}

\centerline{Comment on}

\centerline{
Anyon in an external electromagnetic field:
Hamiltonian and Lagrangian formulations}

\centerline{M.~Chaichian, R.~Gonzales Felipe and D.~Louis Martinez}

\centerline{Phys.~Rev.~Lett. {\bf 71}, 3405 (1993)}

\smallskip
\centerline{by}
\smallskip
\centerline{R.~Jackiw (MIT) and V.~P.~Nair (C.C.N.Y.)}

\bigskip

In their recent letter, Chaichian {\it et.~al.}\ (CFM) present a Lagrangian
for a massive ($m$) point particle on the plane,
with which they claim to realize
anyon statistics \cite{1}.  However, we find that there are some inaccuracies
in their
formulation and, when these are taken care of,
well-known results are reproduced,
which already exist in the literature \cite{2}.
The CFM Lagrangian is
$L_{\rm CFM} = m \left( \dot{x} \cdot \dot{n} \right) / \sqrt{\dot{n}^2}$.
The particle coordinate is
$x^\mu(\tau)$; $n^\mu(\tau)$ is an auxiliary variable for which it is asserted
that $n^2 = -1$ (space-like), but this constraint is not enforced by
$L_{\rm CFM}$,
rather it is imposed ``by hand''.  However, the fixed magnitude of $n$ renders
inconsistent
the equations
implied by
$L_{\rm CFM}$.
These equations read
$$
\eqalignno{
p_{\mu} &\equiv
{\partial L_{\rm CFM} \over \partial \dot{x}^\mu}
= m \dot{n}_\mu / \sqrt{\dot{n}^2} ~~,
{}~~~~~ \dot{p}_\mu = 0
& \hbox{(1a,b)} \cr
p_\mu n^\mu &= 0 ~~, ~~~~~ p^2 = m^2
& \hbox{(2a,b)} \cr
p_{\mu}^{(n)} &\equiv {\partial L_{\rm CFM} \over \partial \dot{n}^\mu}
= {m \over \sqrt{\dot{n}^2}}
\left\{ \dot{x}_\mu - \dot{n}_\mu
{ \dot{x} \cdot \dot{n} \over \dot{n}^2} \right\}
& {(3)} \cr
p^\mu \cdot p_\mu^{(n)} &= 0 &{(4)}}
$$
But from (1) it follows by virtue of (2) that
$0 = n^\mu \dot{p}_\mu = -\dot{n}^\mu p_\mu = -m \sqrt{\dot{n}^2}$.
Vanishing $\dot{n}^2$ contradicts the authors' statement
``assume $\dot{n}^\mu$ to be a timelike vector''
and renders (1a) and (3) meaningless.  Indeed it is easy to
show that the above equations imply $n^\mu$ is constant,
$$
\dot{n}^\mu = 0
\eqno{(5)}
$$
which accentuates the problem.

While
$L_{\rm CFM}$, which is quadratic in derivatives, is inconsistent, a
formulation linear in derivatives ({\it e.g.} as advocated in \cite{3}) may be
given.  The alternative Lagrangian is taken to be
$$
\tilde{L}_{\rm CFM}
= -p_\mu \dot{x}^\mu - p_\mu^{(n)} \dot{n}^\mu - {\lambda \over 2}
(n^2 + 1) - \Lambda_1(p^2 - m^2) - \lambda_1 (p_\mu n^\mu)
- \lambda_2 (p^\mu p_\mu^{(n)})
\eqno{(6)}
$$
The condition $n^2 = -1$ is enforced explicitly.  Quantization proceeds by
recognizing that the last two constraints are second-class (in Dirac's
terminology).  We incorporate them through Dirac brackets, thereby eliminating
them.  The first two constraints are first-class and are retained as
conditions on states.  This leads to the following non-vanishing commutators:
$$
\eqalignno{
[x_{\mu}, \, x_\nu]
&= i \left( n_\mu p_\nu^{(n)} - n_\nu p_\mu^{(n)} \right) / p^2
& \hbox{(7a)} \cr
[x_{\mu}, \, p_\nu]
&= - i g_{\mu\nu} & \hbox{(7b)} \cr
[x_{\mu}, \, n_\nu]
&= i n_\mu p_\nu  / p^2 & \hbox{(7c)} \cr
[x_{\mu}, \, p_\nu^{(n)}]
&= i p_\mu^{(n)} p_\nu  / p^2 & \hbox{(7d)} \cr
[n_{\mu}, \, p_\nu^{(n)}]
&= - i \left( g_{\mu\nu} - p_\mu p_\nu / p^2 \right) & \hbox{(7e)} }
$$
[Except for (7e), these coincide with the Dirac brackets (11) in [1].]
The Hamiltonian reads $H={\lambda \over 2} (n^2 +1) + \Lambda_1 (p^2 - m^2 )$
and generates the equations $\dot{p}_\mu = 0$ and $\dot{n}^\mu = 0$,
consistent with (1) and (5).  Also one gets
$\dot{x}_\mu = -2\Lambda_1 p_\mu , ~ \dot{p}_\mu^{(n)} = \lambda n_\mu$.
These may be integrated to
$$
\eqalignno{
x^\mu(\tau) &= \alpha (\tau) p^\mu + x^\mu_0  &(8) \cr
p_\mu^{(n)} (\tau) &= \beta(\tau) \eta_\mu + p_\mu^{0 \, (n)} &(9)
}
$$
with $\dot{\alpha} = -2\Lambda_1$,
$\dot{\beta} = \lambda$,
and $p^2 = m^2$,
$n^2 = -1$,
$p \cdot n = 0$,
$p \cdot p^{0 \, (n)} = 0$,
the last two implying $p \cdot p^{(n)} = 0$.
In this way the problematic Eqs.~(1a) and (3) are avoided.  Evidently the
motion of {\bf x} is free:
${\bf x}(t) = {\bf v}t + {\tenrm constant}$,
where ${\bf v} \equiv {\bf p} / \sqrt{{\bf p}^2 + m^2}$ is constant.

The anomalous spin emerges when we define (following CFM)
$$
\eqalignno{
n_\mu p_\nu^{(n)} - n_\nu p_\mu^{(n)} &= - \epsilon_{\mu\nu\alpha} S^\alpha
& \hbox{(10a)} \cr
S^\alpha &= -s p^\alpha / (p^2)^{1\over2} & \hbox{(10b)} }
$$
That $S^\alpha$ is proportional to $p^\alpha$ follows from the constraints
{\it i.e.} from the fact that the left side of (10a) is transverse to $p$.
Also since
$s=({p^2})^{-{1\over 2}} \epsilon^{\alpha\beta\gamma} n_\alpha p_\beta^{(n)}
p_\gamma$
it is seen that $s$ is conserved and commutes with $(x^\mu,~p_\nu)$.
Moreover, the Dirac commutators (7) imply
that $S^\alpha$ commutes with itself, as can be verified either by using
the left side of (10a) or the right side of (10b).  [This
contradicts Eq.~(8) in \cite{1}.]  However upon forming the combination
$$
J^\alpha = -\epsilon^{\alpha\mu\nu} x_\mu p_\nu + S^\alpha
\eqno(11)
$$
one establishes the $O(2,1)$ Lorentz algebra,
which would not be satisfied by $-\epsilon^{\alpha\mu\nu} x_\mu p_\nu$ alone.
$$
[J^\alpha, J^\beta] = i \epsilon^{\alpha\beta\gamma} J_\gamma \eqno(12)$$

Thus $S^\alpha$ does indeed describe an anomalous spin.  But now it is
recognized that this realization, when viewed solely in the $x$-$p$ subspace
of variables, {\it i.e.\/}\ using
$$
\eqalignno{
[x_\mu, \, x_\nu ] &= i \epsilon_{\mu\nu\alpha}
{s p^\alpha \over (p^2)^{3\over2}} & \hbox{(13)}\cr}
$$
and (7b), (10b), (11), (12),
is nothing but a ``magnetic monopole'', whose strength is not quantized,
because its space is non-compact $p$-space.  Dynamics can be described solely
in terms of $x$ and $p$ based on the Lagrangian
$$
L \, d\tau = -p_\mu dx^\mu + A_\mu(p) dp^\mu - {\Lambda}_1
(p^2 - m^2) d\tau
\eqno{(14)}
$$
where $A \equiv A_\mu(p) dp^\mu$ is the Dirac-monopole 1-form, whose 2-form
$dA =
{s \over {2(p^2)^{3/2}}} \epsilon_{\alpha\mu\nu} p^\alpha dp^\mu dp^\nu$
(wedge or antisymmetrized product for differential forms is understood).
Also, we see that the motion is free.
$$
\dot{p}_\mu = 0 ~~,~~~~~ \dot{x}^\mu = - 2 \Lambda_1 p^\mu
\eqno{(15)}
$$
The Lagrangian (14) leads to the symplectic structure
$$
\omega ~=~ -dp_\mu dx^\mu ~+~
{s\over {2(p^2)^{3\over 2}}}\epsilon_{\alpha \beta \gamma}
p^\alpha dp^\beta dp^\gamma \eqno(16)
$$
and the consequent commutation rules (13).
This realization of anomalous spin is well known \cite{2}.

The crucial ingredients for a
description of anyons are thus seen to be
the symplectic 2-form $\omega$ of (16) and the mass-shell
condition. Thus seemingly different Lagrangians
are possible, provided they enforce
(16) and the mass-shell condition.
An alternative to (14), which is in fact a different
parametrization of the dynamical variables, is \cite{4}
$$
\tilde{L} \, d\tau ~=~
-p_\mu {dx^\mu} ~-~ {is\over 2} \, {\rm Tr} \,
(\Lambda^{-1}{d \Lambda} \, M_{12})
\eqno(17)
$$
where $\Lambda$ is an $SO(2,1)$-matrix and $M_{12}$ is the generator of the
rotation subgroup $Rot$ of $SO(2,1)$. $\Lambda$ obeys the constraint equation
$$
\Lambda_0^{\mu}= {p^\mu /{(p^2)^{1\over 2}}}, ~~~~{\rm or}~~~~
\eta^{\nu}\Lambda_\nu^\mu  ~=~ {p^\mu /(p^2)^{1\over 2}},~~~~~~~
\eta^\nu ~= (1,0,0) \eqno(18)
$$
The Lagrangian (17) leads to the symplectic structure
$$
\omega ~=~ -dp_\mu dx^\mu ~+~ {is\over 2} {\rm Tr}(\Lambda^{-1} d\Lambda
{}~\Lambda^{-1}d \Lambda ~
M_{12}) \eqno(19)
$$
This is equivalent to (16) by virtue of the constraint (18).
The parametrization in
(18) shows that the part of the symplectic form which describes
spin is the Kirillov-Kostant form on $SO(2,1)/Rot$.

\smallskip
\hbox to \hsize{\hrulefill}
\medskip

\item{1.~}
M.~Chaichian, R.~Gonzales Felipe and D.~Louis Martinez,
{\it Phys.~Rev.~Lett.}~{\bf 71}, 3405 (1993).

\item{2.~}
R.~Jackiw and V.~P.~Nair,
{\it Phys.~Rev.~D}~{\bf 43}, 1933 (1991);
S.~Forte,
{\it Rev.~Mod.~Phys.}~{\bf 64}, 192 (1992).

\item{3.~}
L.~Faddeev and R.~Jackiw,
{\it Phys.~Rev.~Lett.\/}~{\bf 60}, 1692 (1988).

\item{4.~}
A.P.~Balachandran {\it et al},
{\it Gauge~ Symmetries~ and ~Fibre ~Bundles\/},
(Springer, Berlin, 1983) and references therein; S.~Forte, {\it Int.
{}~J.~Mod.~Phys.},
{\bf A7}, 1025 (1992).

\bye